\documentclass[pra,onecolumn,12pt]{revtex4}
\usepackage{amsmath}
\usepackage{graphicx}
\usepackage{bm}
\usepackage{bbold}
\usepackage[dvips]{color}

\begin{document}

\title{Toward Spin Squeezing with Trapped Ions}


\author{Hermann Uys,$^{1}$\protect\footnote{Electronic address:\textit{huys@csir.co.za}},
Michael Biercuk$^{2}$\protect\footnote{Electronic address:\textit{biercuk@physics.usyd.edu.au}}, 
Joe Britton$^{3}$\protect\footnote{Electronic address:\textit{joe.britton@gmail.com}},
John J. Bollinger$^{3}$\protect\footnote{Electronic address:\textit{john.bollinger@boulder.nist.gov}}
}

\affiliation{\mbox{$^1$ National Laser Centre, Council for Scientific and Industrial Research, Pretoria, South Africa}\\
\mbox{$^2$ School of Physics, University of Sydney, Sydney, Australia}\\
\mbox{$^3$ National Institute of Standards and Technology, Boulder, Colorado, USA} }

\begin{abstract}
Building robust instruments capable of making interferometric measurements with precision beyond the standard quantum
limit remains an important goal in many metrology laboratories.  We describe here the basic concepts underlying spin
squeezing experiments that allow one to surpass this limit.  In priniciple it is possible to reach the so-called
Heisenberg limit, which constitutes an improvement in precision by a factor $\sqrt{N}$, where $N$ is the number of
particles on which the measurement is carried out. In particular, we focus on recent progress toward implementing spin
squeezing with a cloud
of beryllium ions in a Penning ion trap, via the geometric phase gate used more commonly for performing two-qubit
entangling operations in quantum computing experiments.  
\end{abstract}

\maketitle


\section{Introduction}

Exploitation of coherent quantum phenomena represents a new frontier in the field of metrology, that study which aims to
achieve measurements of physical phenomena with ever increasing precision.  Probably \textit{the} prototypical quantum
metrology experiment is the simple Ramsey interferometry measurement used in atomic clocks, which for decades has
been the basis for the calibration of time and frequency standards.
Modern quantum metrology experiments, however, often entail sophisticated manipulation of several quantum degrees of
freedom to achieve a single
measurement outcome.  As an example consider the quantum-logic spectroscopy clock measurements in which the state of an
internal clock transition of one atom is transferred to a detectable transition in an auxiliary atom using the quantum
mechanical motion of the atoms as a bus \cite{Chou2010}.

A natural limit in precision of a measurement carried out on an ensemble of $N$ uncorrelated particles is the standard
quantum limit, in which the measurement precision scales with $\sim1/\sqrt{N}$.  This scaling is a direct consequence
of the Poissonnian statistics resulting from the lack of correlation between the measurement outcomes of the individual
particles.  On the other hand, the intrinsic limit of a quantum measurement is governed by Heisenberg's uncertainty
principle and allows the precision to scale as $\sim1/N$, the Heisenberg limit \cite{CombesWiseman2005}.  Any quantum
state that allows a
measurement precision surpassing the standard quantum limit is said to be ``squeezed''. This proceedings reviews the
mathematical description of squeezing in pseudo-spin systems and current progress toward implementing squeezing in
trapped ion systems.

Several strategies to achieve spin squeezing exist, and a number have been implemented experimentally.  One 
method is through the use of quantum non-demolition measurements.  A typical implementation
relies on the coupling between the Stokes vector of a probe beam of light and the spin state of an ensemble of cold
atoms.  This coupling induces a spin-dependent polarization rotation of the probe beam, which can be measured with a
polarimeter down-stream of the atoms.  The outcome of such a measurement allows one to improve one's prediction for
the outcome of a subsequent measurement, which constitutes a reduction in spin noise, i.e. squeezing of the spin
uncertainty.  Although this measurement conveys information regarding the spin-state it is non-projective and hence is
labelled as a non-demolition measurement.  Oftentimes the amount of squeezing is characterized by comparing the ratio
of the variances of the measured spin component of the squeezed versus unsqueezed states:
\begin{equation}
 \mathcal{S}=10\log{\left(\frac{\sigma^2_{squeeze}}{\sigma^2_{unsqueeze}}\right)}.
\end{equation}
This approach has been used by Appel \textit{et al.} \cite{Appel2009} as well
as Takano \textit{et al.} \cite{Takano2009} to produce respectively $\mathcal{S}=-3.4$ dB and $\mathcal{S}=-1.8$ dB of
squeezing.

An alternative approach to squeezing is through engineering of a phase shift that depends nonlinearly on the spin-state
of the particles.  In cold, neutral atomic gases one way to realize such a nonlinearity is via the mean-field
interaction resulting from interparticle s-wave scattering.  This interaction is ever-present, so the technical
challenge is to
turn it on, or off, controllably.  This ability was recently demonstrated by two groups.  Riedel \textit{et al.}
\cite{Riedel2010}~studied a
two-component Bose-Einstein condensate formed by atoms in different hyperfine states.  They controlled the
nonlinearity by employing a trap which can indepently manipulate the trapping potential of the two components.  Since
the mean-field interaction depends on the overlap of the wavefunctions of the two components, which their trap
allowed them to vary, they were able to
controllably squeeze and achieve a reduction in spin noise of $\mathcal{S}=-3.7$ dB. Gross and co-workers
\cite{Gross2010} achieved the same goal by instead tuning the s-wave scattering length using a narrow Feschbach
resonance, obtaining $\mathcal{S}=-8.2$ dB of squeezing.  

A nonlinearity can also be engineered by placing an ensemble of atoms in an optical cavity.  The presence of light
circulating in such a cavity induces an ac Stark shift on the hyperfine groundstate levels close to resonance with the
cavity mode.  The atoms in turn modify the index of refraction in the cavity which shifts the cavity resonance
frequency.  As a result, the light shift on each atom depends on the presence of all other atoms in the cavity leading
to the nonlinear phase shift required for squeezing.   Leroux and others \cite{Leroux2010} achieved
$\mathcal{S}=-5.6$dB of squeezing using this approach.  Recently, using a similar approach Thompson and co-workers
achieved $\mathcal{S}=-3.4$ dB of squeezing on nearly $10^{6}$ Rb atoms \cite{ChenBohnet2011}.

The aim of this proceedings is to give a pedagogic overview of the formalism needed to describe squeezing in
an ensemble of two-level systems, as well as the quantum optics relevant to a particular implementation
using trapped ions in a Penning trap that depends on creation of a nonlinear phase shift. The
rest of this manuscript is organized as follows: in the next section we summarize
the mathematical language required to describe the squeezing effect in an ensemble of two-level particles.
We then consider the spin squeezing interaction first discussed in detail by Kitagawa and Ueda \cite{Kitagawa1993}.
This is followed by a description of how the squeezing interaction can be engineered in a trapped ion system, which
builds on the first squeezing demonstrations with two trapped ions in a radio-frequency Paul trap \cite{MeyerRowe2001}.
The particulars of the Penning ion
trap experiment are then discussed before the final section looks at some of the technical challenges faced in
successfully implementing the experiment.

\section{Pseudo Spin Algebra}

In this section we develop a convenient mathematical description for an ensemble of $N$ two-level systems of which the
two levels are coupled by an oscillating electric or magnetic field.  Assuming each particle can be in either of the
states
$|\!\downarrow\rangle$, $|\!\uparrow\rangle$, the many particle Hamiltonian coupling the two states can be expressed in
a frame rotating at the transition frequency  between the levels as
\begin{equation}
H = \sum\limits_{i=1}^N\frac{\hbar}{2}\delta\sigma^z_i +
\hbar\Omega_R\sum\limits_{i=1}^N\left(\sigma_i^- + \sigma_i^+\right),\label{Hmany}
\end{equation}
where the index $i$ labels the $i$'th particle.
Here $\delta=\omega_0-\omega$ is the detuning between the transition frequency $\omega_0$ and drive field frequency
$\omega$, $\sigma^z_i$ is the Pauli operator, obeying the usual spin angular momentum commutation
relation $\left[\sigma^x_i,\sigma^y_i\right]=2i\varepsilon_{xyz}\sigma^z_i$, while the raising and lowering operators
are defined via $\sigma_i^+=\frac{1}{2}(\sigma_i^x+i\sigma_i^y)$, $\sigma_i^-=\frac{1}{2}(\sigma_i^x-i\sigma_i^y)$ and
have the effects: $\sigma^+_i|\!\downarrow\rangle=|\!\uparrow\rangle$, 
$\sigma^-_i|\!\uparrow\rangle=|\!\downarrow\rangle$.
It is convenient to  introduce the following pseudo-spin operators
\begin{equation}
J_z=\sum\limits_i\frac{1}{2}\sigma^z_i,\hspace*{1cm} 
J_+=\sum\limits_i\sigma^+_i,\hspace*{1cm}J_-=\sum\limits_i\sigma^-_i.
\end{equation}
This transformation preserves the spin commutator relation $\left[J_x,J_y\right]=2i\varepsilon_{xyz}J_z$ and allows the
the Hamiltonian to be written concisely as
\begin{equation}
H = \hbar\delta J^z + \hbar\Omega_R\left(J^- + J^+\right).\label{HJ}
\end{equation}
As typical experiments are initiated with all particles optically pumped to a specific state, we consider the action
of the lowering operator on that initial state with all particles in $|\!\uparrow\rangle$, i.e. $|J=N/2,
M_J=N/2\rangle=|N/2,
N/2\rangle=|\uparrow\uparrow\uparrow...\uparrow\uparrow\rangle$.  The choice of labelling in the left-hand ket in the
latter will become clear presently.  Operating consecutively with the lowering operator we get (the normalization
factors in what follows are explicitly calculated in Appendix A):
\begin{eqnarray}
&&\hspace*{-1.2cm}|N/2,
N/2-1\rangle=\frac{1}{\sqrt{N}}J_-|\uparrow\uparrow\uparrow...\uparrow\uparrow\rangle
=\sum\limits_j\frac{\sigma^-_j}{\sqrt{N}}|\uparrow\uparrow\uparrow...\uparrow\uparrow\rangle=\frac{1}{\sqrt{N}}
\sum\limits_{j}
|\uparrow\uparrow...\downarrow_j...\uparrow\uparrow\rangle\\
&&\hspace*{-1.2cm}|N/2, N/2-2\rangle=
\frac{1}{\sqrt{2!N(N-1)}}J_-^2|\uparrow\uparrow\uparrow...\uparrow\uparrow\rangle\\
&&\hspace*{1.4cm}=\frac{1}{\sqrt{2!N(N-1)}}\sum\limits_{j_1,j_2}
\sigma^-_{j_1}\sigma^-_{j_2}|\uparrow\uparrow\uparrow...\uparrow\uparrow\rangle\\
&&\hspace*{1.4cm}=\frac{\sqrt{2!}}{\sqrt{N(N-1)}}\sum\limits_{j_1<
j_2}|\uparrow\uparrow...\downarrow_{j_1}...\downarrow_{j_2}\uparrow\rangle
\label{J2state}
\end{eqnarray}
The state on the right hand side in \eqref{J2state} is therefore a symmetric superposition of all number states for
which two particles are in the down state and all other particles in the up state.  Likewise after applying $J_-$
for $N/2$ times, one gets
\begin{eqnarray}
|N/2,0\rangle\!\!\!\!\!&=&\!\!\!\!\!\frac{1}{\sqrt{N!}}J_-^{N/2}|\uparrow\uparrow\uparrow...\uparrow\uparrow\rangle\\
\!\!\!\!\!&=&\!\!\!\!\!\frac{1}{\sqrt{N!}}\sum\limits_{j_1,j_2,...j_{N/2}}\sigma^-_{j_1}\sigma^-_{j_2}...\sigma^-_{j_{
N/2}} |\uparrow\uparrow\uparrow...\uparrow\uparrow\rangle\\
\!\!\!\!\!&=&\!\!\!\!\!\frac{(N/2)!}{\sqrt{N!}}\sum\limits_{j_1<j_2<...<j_{N/2}}|\uparrow...\downarrow_{j_1}
..\downarrow_{j_2} ...\downarrow_ { j_ { N/2}} ..\uparrow\rangle ,
\end{eqnarray}
and $N$ times
\begin{equation}
|N/2,-N/2\rangle=J_-^{N}
|\uparrow\uparrow\uparrow...\uparrow\uparrow\rangle=|\downarrow\downarrow\downarrow...\downarrow\downarrow\rangle.
\end{equation}
Or in general
\begin{equation}
|N/2,N/2-q\rangle=\frac{\sqrt{q!(N-q)!}}{\sqrt{N!}}\sum\limits_{j_1<j_2<...<j_{q}}|\uparrow...
\downarrow_{j_1}..\downarrow_{j_2} ...\downarrow_ { j_ {q}} ..\uparrow\rangle.
\end{equation}
It is now apparent that the second index in our labelling of the left-hand kets indicates one-half of the difference
between the number of particles in the state $|\!\uparrow\rangle$ versus $|\!\downarrow\rangle$.  That number is exactly
the eigenvalue of the $J_z$ operator, as is clear from the following examples:
\begin{eqnarray}
&&\hspace*{-0.8cm}J_z|\uparrow\uparrow\uparrow...\uparrow\uparrow\rangle
=\sum\limits_i\frac{1}{2}\sigma^z_i|\uparrow\uparrow\uparrow...\uparrow\uparrow\rangle=\frac{N}{2}
|\uparrow\uparrow\uparrow...\uparrow\uparrow\rangle\\
&&\hspace*{-0.8cm}J_z(\frac{1}{\sqrt{N}}J_-|\uparrow\uparrow\uparrow...\uparrow\uparrow\rangle)
=\frac{1}{\sqrt{N}}\sum\limits_{i,j}\frac{1}{2}\sigma^z_i|\uparrow\uparrow...\downarrow_j..\uparrow\uparrow\rangle\\
&&\hspace*{-0.8cm}\hspace*{3.9cm}=(\frac{N}{2}-1)\frac{1}{\sqrt{N}}\sum\limits_i|\uparrow\uparrow..
\downarrow_i...\uparrow\uparrow\rangle\\
&&\hspace*{-0.8cm}J_z\frac{1}{\sqrt{2!N(N-1)}}J_-^2|\uparrow\uparrow...
\uparrow\uparrow\rangle=(\frac{N}{2}-2)\frac{1}{\sqrt{2!N(N-1)}}\sum\limits_{j_1\neq j_2}|\uparrow...
\downarrow_{j_1}...\downarrow_{j_2}\uparrow\rangle.
\end{eqnarray}
We have thus constructed a set of basis states that are eigenstates of the $J_z$ operator, i.e. $J_z|N/2,N/2-m\rangle =
(N/2-m)|N/2,N/2-m\rangle$ with $m=0,1,2,...N$.  With some more algebra it can be shown that these states are
simultaneously eigenstates of the operator $J^2=J^2_x+J^2_y+J^2_z$, with the degenerate eigenvalue $(N/2)(N/2+1)$.  This
motivates the choice of labelling for the first index.  The set of states $|N/2,N/2-m\rangle$
therefore behave as eigenstates of a spin angular momentum with magnitude $N/2$, and can be denoted as
$|J,M_J\rangle$. We will refer to these as a
\textit{pseudo-spin}.  These states are well known from the theory of superradiance described by Dicke in 1954
\cite{Dicke1954}
and are often referred to as Dicke states. It is important to note that the states do not span the entire Hilbert space,
but just the particular symmetric subspace constructed here.  However, since we will consider only experiments initiated
with all particles pumped into the state $|\!\uparrow\rangle$, the dynamics accessible through \eqref{HJ} restricts the
system to this symmetric subspace. \\

\begin{figure}
  \includegraphics[height=.3\textheight]{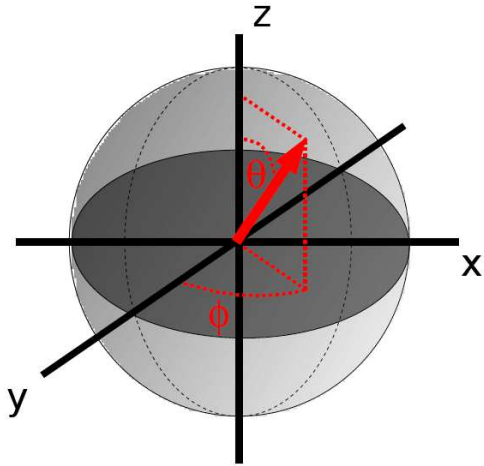}
\label{blochvec}
  \caption{The state of an $N$-particle pseudo-spin system can be represented by a Bloch-vector vector of length
$J=N/2$.}
\end{figure}

A useful pictorial description of a single two-level system can be brought to bear here,
namely the concept of a Bloch vector.  Any state of a single two-level system can be represented by the state vector
with coefficients $(e^{-i\phi}\cos{\theta}, e^{i\phi}\sin{\theta})$.  One can think of the parameters as being
the polar angle, $\theta$, and the azimuthal angle, $\phi$, of a unit vector in 3D space, called  the Bloch vector as
shown in figure \ref{blochvec}. Likewise, we can represent the state of the $N$-particle system by a vector of length
$N/2$. In analogy to the single particle case, when the Bloch vector points along the positive $z$-axis it represents a
state with all particles in $|\!\uparrow\rangle$, when it points along the negative $z$-axis it represents a state
with all particles in $|\!\downarrow\rangle$.  As a result of the uncertainty relation $\Delta J_x\Delta J_y\geq \langle
J_z/4\rangle$ the projections of the Bloch vector onto the $x$ and $y$ axis when in the
state $|N/2,N/2\rangle$ are non-zero, so that the vector should instead of a line element, be thought of as a cone with
its apex at the origin, and the width of which indicates the uncertainty in the projections.  

To complete the picture for other directions on the Bloch sphere we have to consider the dynamics due to \eqref{HJ},
which for $\delta=0$ can be simply rewritten as $H=\hbar\Omega_R\hat J_x$. The time evolution then becomes $U(t) =
\exp{(-i\Omega_R\hat J_xt)}$ which one recognizes as the rotation operator for the pseudo-spin which causes a rotation
of
the vector through an angle $\Omega_Rt$ around the $x$-axis.  Starting with the state $|N/2,N/2\rangle$ and
operating with $\hat U(\Omega_Rt=\pi/2)$ therefore places the Bloch vector along the $y$-axis.  By expanding the
$\hat U(\pi/2)$ one finds that the state of the system is
\begin{equation}
 |CS\rangle = \frac{1}{2^{N/2}}\sum\limits_{M_J=-N/2}^{N/2}\begin{pmatrix}N\\N/2+M_J\end{pmatrix}^{1/2}|N/2,
M_J\rangle\label{CS}
\end{equation}
which is known as a coherent spin state.  The uncertainty in a measurement of $\langle CS|J_z|CS\rangle$ scales as
$1/\sqrt{N}$, i.e. the standard quantum limit, due to the binomial distribution of amplitudes in \eqref{CS}.

\section{Kitagawa Shearing Gate}

\begin{figure}
\hspace*{-1.2cm} \includegraphics[height=.22\textheight]{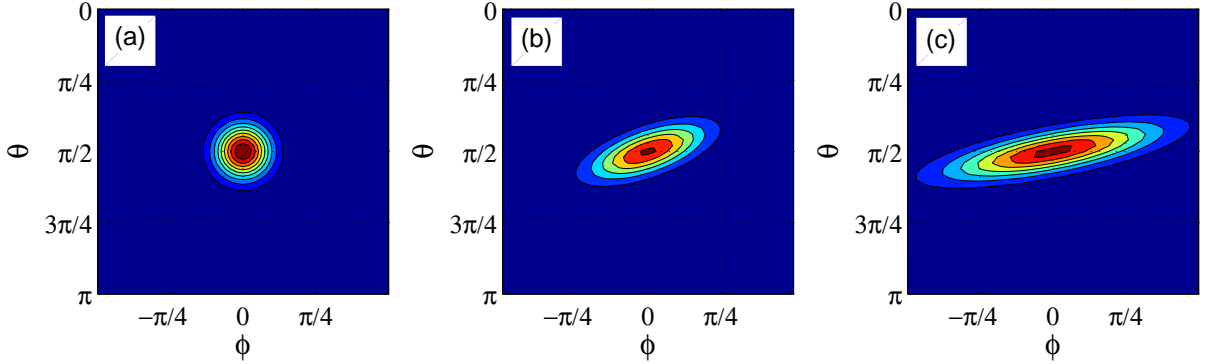}
  \caption{Probability of projecting the squeezed state $\hat U_{Sq}(\chi t)|CS\rangle$ on the rotated state
$\exp(-i\phi\hat J_z)\exp(-i\theta\hat J_x)|CS\rangle$ where the squeezing operator was applied for times (a) $\chi t =
0$, (b) $\chi t = 0.05$, and (c) $\chi t = 0.1$, respectively.}
\end{figure}

\begin{figure}
  \includegraphics[height=.3\textheight]{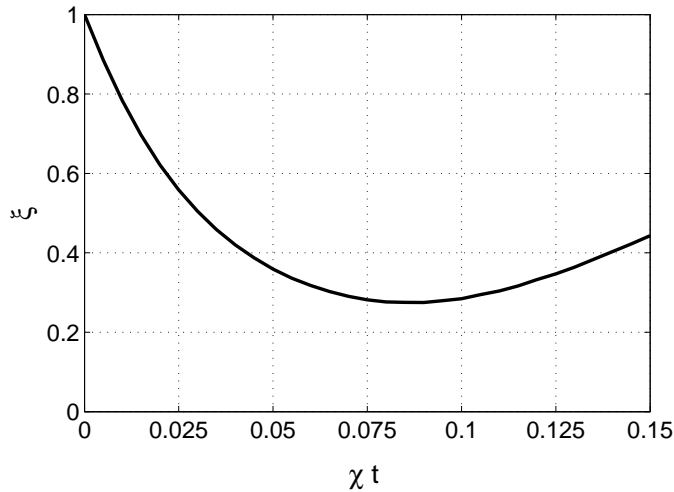}
\label{xivt}
  \caption{The squeezing parameter $\xi=\Delta J_\perp/\sqrt{J/2}$ as a function of time.}
\end{figure}

In a seminal paper in 1993 \cite{Kitagawa1993} Kitagawa and Ueda described spin-squeezing in pseudo-spin systems which
results from
unitary transformations of the form 
\begin{equation}
\hat U_{Sq}(t)=\exp{\left[-i\chi \hat J^2_zt\right]}.\label{squeezeop}
\end{equation}
The effect of the squeezing operator $\hat U_{Sq}(t)$ on any superposition of spin states is to induce a phase shift
that depends nonlinearly on the $\hat J_z$ eigenvalue of each state.  We consider now the resulting dynamics on the
coherent
state $|CS\rangle$ defined above.  Figure \ref{conefig} shows a sequence of snapshots of the uncertainty distribution
of $|CS\rangle$ for $J=25$ after it has been subjected to $\hat U_{Sq}(t)$ for times (a) $\chi t = 0$, (b) $\chi t =
0.05$ and (c) $\chi t = 0.1$ respectively.  As a measure of that uncertainty figure \ref{conefig} plots the
modulus squared of the projection of the state $\hat U_{Sq}(t)|CS\rangle$ onto a rotated coherent state,
$\exp(-i\phi\hat J_z)\exp(-i\theta\hat J_x)|CS\rangle$, as a function of the two rotation angles. It is clear that the
squeezing operator shears the uncertainty cone, reducing the uncertainty along one spin axis at the cost of increasing
the uncertainty along an orthogonal axis.  The state is said to now be ``squeezed''. Kitegawa and Ueda referred to this
effect as one-axis twisting of the
uncertainty cone. Note that as the squeezing takes place, the Bloch vector also shortens even though the system remains
in a pure state. An appropriate parameter to
quantify the metrologically relevant squeezing, which takes into account this shortening, is $\xi=\Delta
J_\perp/\sqrt{J/2}$ where $\Delta J_\perp$ is the
root-mean-squared deviation along the direction of minimum uncertainty. Figure \ref{xivt} plots $\xi$ as a
function of the time that the squeezing operator is applied, again for $J=25$.  The squeezing does not indefinitely
reduce the uncertainty in $J_z$, but eventually reaches a minimum uncertainty  around $\xi=0.28$ for $J=25$. The minimum
value of $\xi$ decreases for larger numbers of spins as the curvature of the Bloch sphere becomes less
\cite{Kitagawa1993}.

It is this single-axis twisting Hamiltonian that we aim to realize by use of trapped ions, as discussed next.

\section{Shearing Gates with Ions}

One way of implementing the squeezing operator \eqref{squeezeop} in trapped ions relies on a generalization of the
geometric ion-qubit phase gate demonstrated by Leibfried \textit{et al.} \cite{Leibfried2003}.  In that experiment the
quantum motion
of ions in a trap is used as a bus to mediate the nonlinear phase shift required to affect squeezing.  An
experimental sequence is as follows:\\  
(1) $N$ ions are trapped and cooled in an ion trap with characteristic center-of-mass frequency $\omega_z$.\\
(2) The internal state of the ions is prepared in the superposition $|CS\rangle$, \eqref{CS}.\\
(3) An oscillatory force is applied to the ions, close to resonance with the axial center-of-mass mode of motion of the
multi-ion system. 
The force is designed to be state dependent, so that ions in the state $|\!\downarrow\rangle$ feel an equal but opposite
force to ions in the state $|\!\uparrow\rangle$. As a result the force will depend on the value of $M_J$ of the
pseudo-spin states.\\
(4) The frequency of the oscillatory force is chosen off-resonant by a detuning $\delta^\prime$, so that after a period 
$t=2\pi/\delta^\prime$ the drive force has completely dephased and rephased with the oscillating ions, thus
accelerating and decelerating it and returning it to its initial motional state.

\begin{figure}
  \includegraphics[height=.3\textheight]{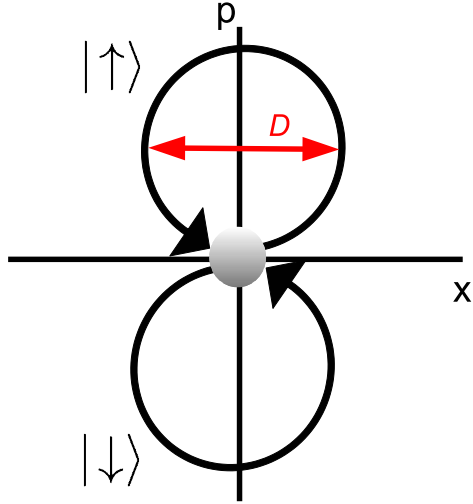}
\label{geomphase}
  \caption{Squeezing is implemented on trapped ions by applying a force on the ions that depends on their internal
state, $|\downarrow\rangle$ or $|\uparrow\rangle$.  This schematic represents an interaction picture in which the time
evolution at the excitation frequency has been removed.  If the ions are excited in such a way that they trace
out closed loops in phase space, as pictured here, they will acquire a geometric phase proportional to the area of the
loop.  That phase has the requisite $\hat J_z^2$ dependence to cause squeezing.}
\end{figure}

A geometric interpretation gives a simple physical picture of how this sequence leads to squeezing \cite{Leibfried2003}.
Since the
ions return to their initial motional state, they trace out a closed loop in phase-space, as pictorially
represented in Fig.~\ref{geomphase}.  As a result, they acquire a geometric phase, $\Phi$, proportional to the area,
$A$, of the loop.
Since the force, $F$, on each ion depends on the internal state of that ion, the force on the multi-ion system depends
on the difference between the number of ions in the states $|\!\downarrow\rangle$ and $|\!\uparrow\rangle$, and
therefore on the eigenvalue of $J_z$ for a state $|N/2,N/2-m\rangle$.  In turn the radius, $R$ of the phase-space loop
depends linearly on the force so that $R\propto F$.  Since the area $A=\pi R^2$ we therefore know that the $\Phi\propto
A\propto J_z^2$. This relation is precisely what is needed to implement \eqref{squeezeop}.\\

\begin{figure}
  \includegraphics[height=.25\textheight]{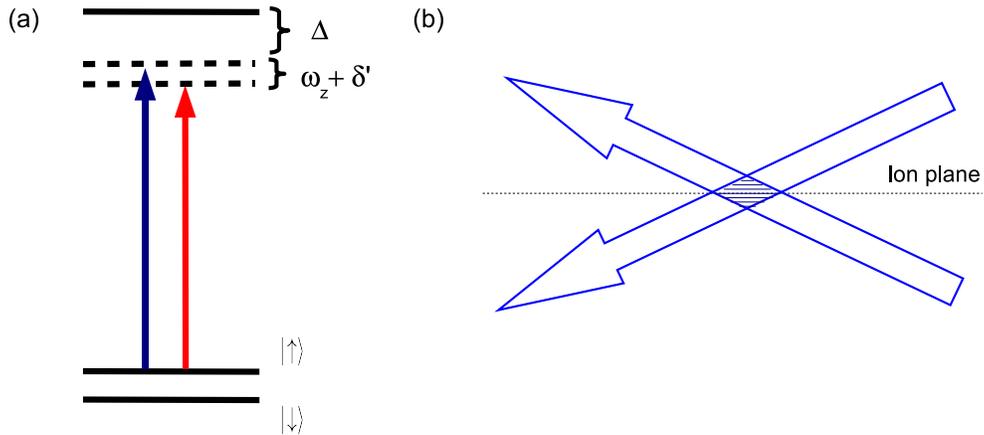}
\label{levels}
  \caption{Experimental scheme for implementing squeezing.  (a) Two laser beams are chosen to have relative detunings
close to the characteristic motional axial center-of-mass frequency, $\omega_z$ of the trapped ions. 
(b) The laser beams are overlapped so as to form a moving interference pattern oscillating at $\omega_z$ and with a wave
vector normal to the plane in which the ions form a planar ion crystal.}
\end{figure}

\begin{figure}
  \includegraphics[height=.3\textheight]{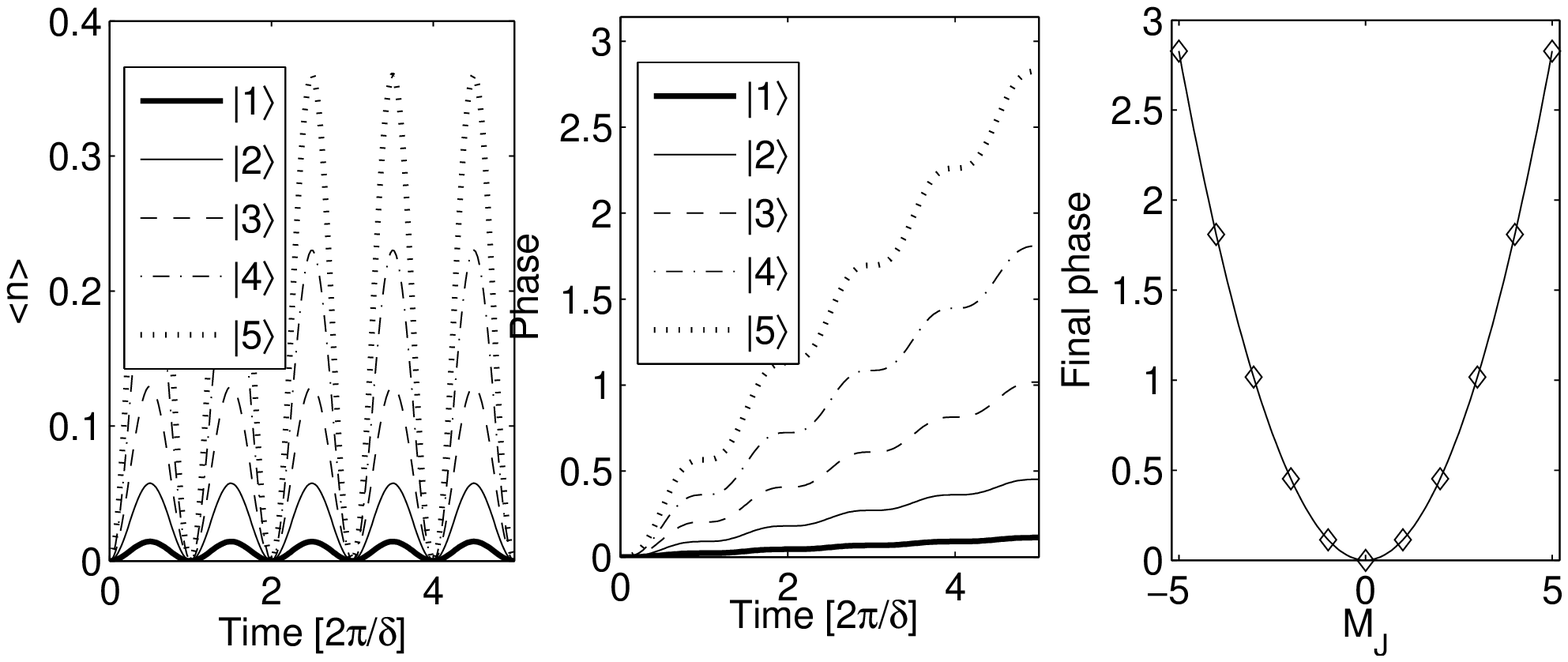}
\label{nbarphase}
  \caption{Dynamics resulting from the Hamiltonian \eqref{Hcoh}.  (a) States starting in the motional ground state, but
with different $M_J$ values, indicated in the legend, are excited and de-excited as is evident from the periodic
increase and decrease in expected excitation number $\langle\hat n\rangle$.  Higher $M_J$ values reach higher excitation
numbers, but all states return to the motional ground state after each period $\chi t=2\pi/\delta^\prime$. (b) The
corresponding phases as a function of time for each state represented in (a).  (c) The parabolic character of the phases
acrued after time $\chi t=5\times 2\pi/\delta^\prime$ illustrates the correct dependence to allow squeezing.
}
\end{figure}

Experimentally the required force is implemented by use of a pair of detuned laser beams illuminating the ions.
Consider an ensemble of ions all lying in a single plane, each with a three-level internal structure, as shown in Fig.
\ref{levels}(a).   The ions are illuminated by the laser beams in a configuration shown in Fig. \ref{levels}(b) and the 
laser beams have frequencies, $\omega_1$ and $\omega_2$. We require that
$\delta^\prime\ll\omega_z\ll\Delta$.  If $\delta_L=\omega_1-\omega_2=0$ the overlapping light
fields will form a standing-wave interference pattern along the direction normal to the plane of the ions.  The
standing wave leads to a position dependent AC Stark shift on the ions, while
the gradient of this shift results in a position-dependent force on the ions.  
Through appropriate choice of laser polarization and detuning, $\Delta$, a different AC Stark shift is obtained for
the $|\!\downarrow\rangle$ and $|\!\uparrow\rangle$ states, resulting in a state-dependent force.
Detuning the light beams by
$\delta_L=\omega_z$ will cause the standing wave to ``walk'', resulting at any point in a periodic force,
oscillating at the characteristic center-of-mass frequency of the trap, which will excite the motion of the ions. 

A key requirement to ensure that the motion gets excited is that the spatial extent of the ions' wavefunction is
significantly smaller than the wavelength of the standing wave.  This condition is referred to as the Lamb-Dicke
criterion and can be expressed as $\eta=x_0k_{eff}\ll1$, where $x_0$ is the characteristic width of the ions' spatial
wavefunction along $\mathbf{k}_{eff}$, the wavevector of the standing wave.  If the Lamb-Dicke criterion is not met, the
ion wavefunction will feel a drive force in one direction at some points and in the opposite direction at others.  The
force will average to zero and no excitation will occur.  

Given that the Lamb-Dicke criterion is fulfilled and $\delta_L=\omega_z+\delta^\prime$, one can write down an effective
Hamiltonian for the interaction between the light fields and the atom as follows:
\begin{equation}
 H = \frac{g^2\eta}{\Delta}(\hat a^\dagger e^{i\delta^\prime t} + \hat ae^{-i\delta^\prime t})J_z.\label{Hcoh}
\end{equation}
Here $g=dE$ where $d$ is the dipole matrix element and $E$ the amplitude of the light field.  \\

Under the action of Hamiltonian \eqref{Hcoh} for a time $2\pi/\delta^\prime$ the ions will trace out precisely the
closed loop in phase space discussed above.  This is illustrated in figure \ref{nbarphase}, where in (a) we plot the
average number of excitations of the center-of-mass oscillator mode, $\langle\hat n\rangle=\langle\hat a^\dagger\hat
a\rangle$, as a function of the time and for different initial states: thick solid line - $|0\rangle|M_J=1\rangle$,
thin solid line - $|0\rangle|M_J=2\rangle$, dashed line - $|0\rangle|M_J=3\rangle$, dot-dashed line -
$|0\rangle|M_J=4\rangle$ and dotted line - $|0\rangle|M_J=5\rangle$.  In the latter the first ket $|0\rangle$
represents the ground state of the motional mode corresponding to the operator $\hat a^\dagger$. As predicted, the
average excitation number
increases and then decreases back to zero for each period $t=2\pi/\delta^\prime$, consistent with the ions undergoing
an excursion in phase space away from its equilibrium position and back.  Moreover, those states with larger $M_J$
undergo larger excursions. Figure
\ref{nbarphase}(b) plots the complex phase as function of time for the same states, demonstrating that this phase
monotonically increases with time and is also dependent on the value of $M_J$.  Figure \ref{nbarphase}(c) plots the
final phase, diamonds, at time $t=5\times2\pi/\delta^\prime$ for all states $|0\rangle|M_J\rangle$ with $-5\leq
M_J\leq5$. The solid line is not a fit, but the function $f(M_J)=5\times4\pi(g^2\eta/\Delta)^2M_J^2/\delta^\prime$,
clearly demonstrating the quadratic dependence of the phase on the value of $M_J$, as expected for the squeezing
operator.

\section{Penning Trap Implementation}

\begin{figure}
  \includegraphics[height=.28\textheight]{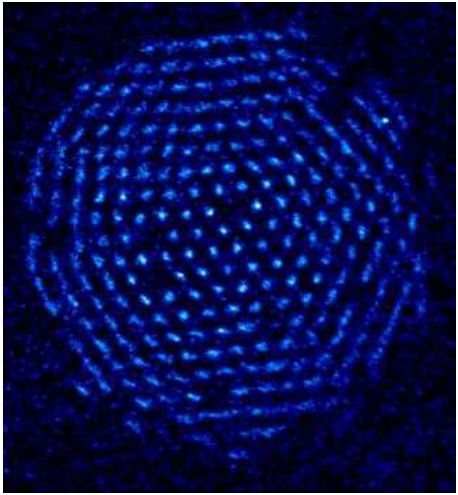}
\label{singleplane}
  \caption{A single plane of ions.  This image was obtained by strobing the camera synchronously with the ion plane
rotation. For details see \cite{MitchellBollinger1998}.}
\end{figure}

Efforts are currently underway at NIST in Boulder, Colorado, USA to implement the scheme discussed above in a Penning
ion trap.  A feature of the Penning trap compared to RF-Paul trap setups is its ability to capture many
ions (hundreds) in a single plane. While a detailed discussion of the plasma physics involved in Penning traps is
beyond the scope of this proceedings we give a brief overview of the aspects most relevant to us.  

To confine ions a Penning trap uses a combination of a static, cylindrically symmetric electric quadrupole field, and a
static magnetic field oriented parallel to the axis of symmetry of the electric field \cite{Biercuk2009}.  The electric
field
provides harmonic axial confinement and the magnetic field confinement in the radial plane in which a single isolated
ion will undergo complicated epitrochoid motion.  A cloud of ions in the trap will rotate rigidly around the symmetry
axis, and the geometry of the cloud depends sensitively on the rotation frequency, varying from cigar-shaped to
pancake-shaped.  When sufficiently cold, and at the appropriate rotation frequency the ions will crystalize into a disc
consisting
of a single ion layer.  Such a disc is pictured in Fig.~\ref{singleplane}.
\begin{figure}
  \includegraphics[height=.3\textheight]{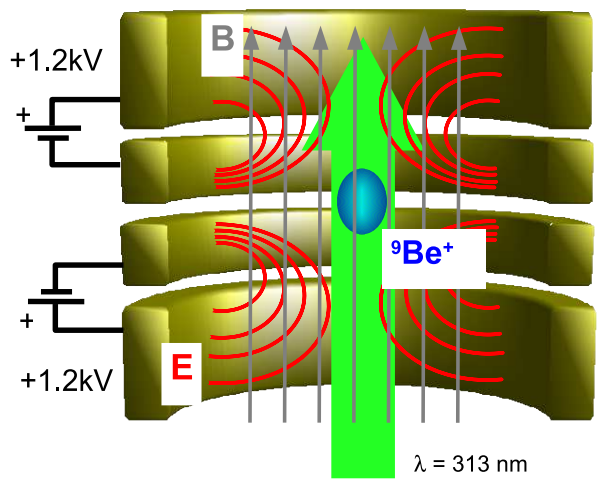}
\label{trap}
  \caption{Design of the NIST Penning ion trap.  Static voltages are applied to four stacked cylindrical electrodes to
provide a quadrupole electric field that confines the ions along the axial direction.  A 4.5 T magnetic field provided
by a superconducting magnet provides confinement in the transverse plane.}
\end{figure}

The physical trap, Figure \ref{trap}, consists of two pairs of stacked cylindrical electrodes onto which static voltages
of 1.2 kV are applied to provide the quadrupole electric field.  A magnetic field of 4.5 T is provided by a
superconducting
magnet. Cooling laser beams enter the trap both axially and perpendicular to the trap axis.  The ions can be imaged both
by a side-view or top-view camera.  Taking stroboscopic images with the top-view camera reveals the crystalline
structure of the ions, see Fig.~\ref{singleplane}.  For a single plane the ions for a triangular lattice.
\begin{figure}
  \includegraphics[height=.4\textheight]{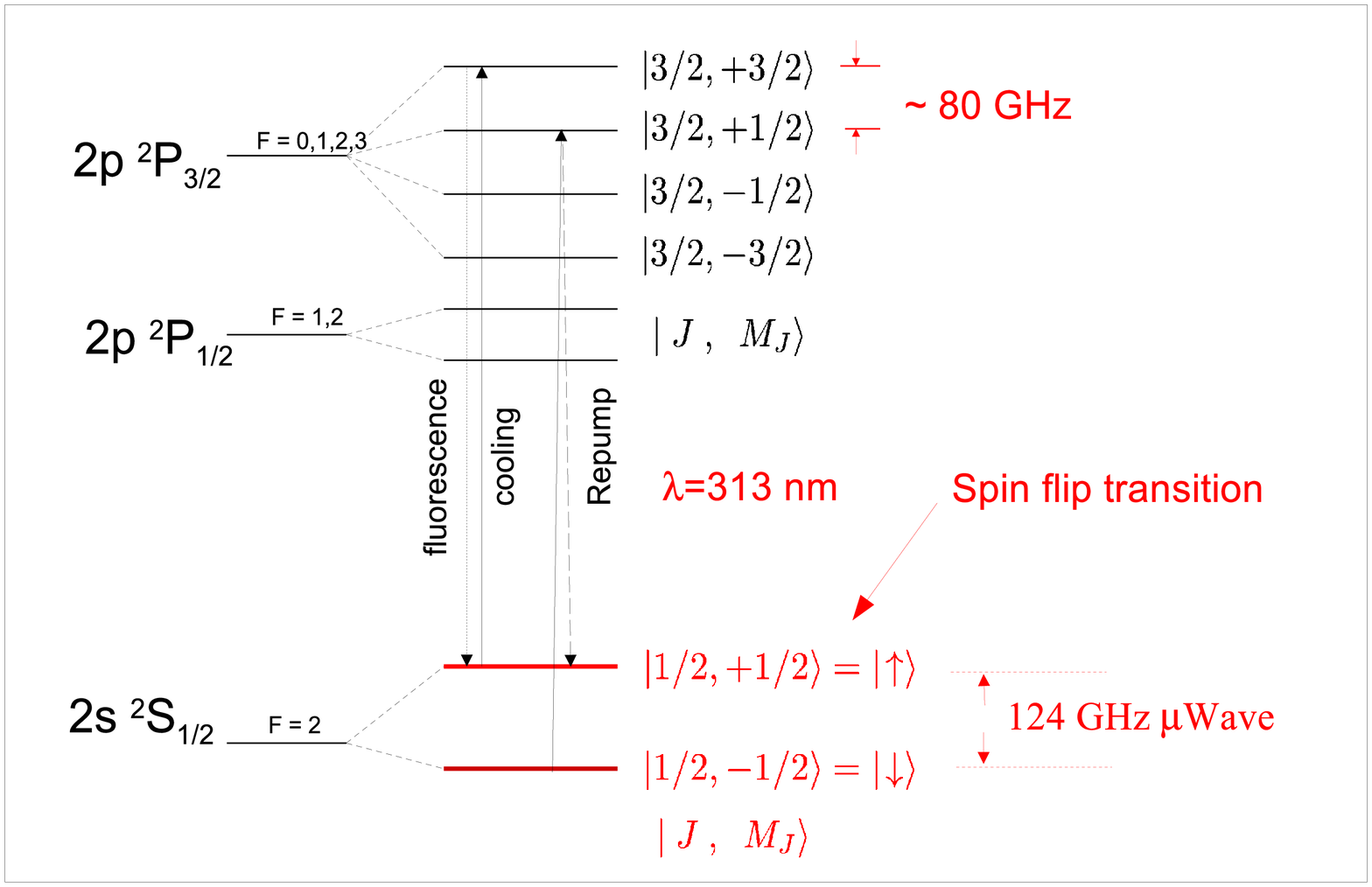}
\label{Belevels}
  \caption{Level structure of $^9$Be$^+$ at 4.5 T.  We neglect the Be$^+$ nuclear spin $(I=3/2)$ in this diagram.  The
nuclear spin is optically pumped to the $M_I=3/2$ state throughout the duration of an experiment.}
\end{figure}

The NIST experiment traps $^9$Be$^+$ ions for which the level structure is given in figure \ref{Belevels}. The states
$|J=1/2,M_J=1/2\rangle$ and $|J=1/2,M_J=-1/2\rangle$ in the $^2$S$_{1/2}$ hyperfine groundstate manifold serve as the
spin states $|\!\uparrow\rangle, |\!\downarrow\rangle$ respectively.  They are separated by 124 GHz and controlled by
microwaves generated by a p-i-n diode oscillator and injected into the trap through a microwave horn.  The microwave
field allows high fidelity rotations on the Bloch-sphere.  Doppler cooling is achieved by off-resonant scattering of
313 nm light between the states $|\!\uparrow\rangle$ and $|J=3/2,M_J=3/2\rangle$ in the $^2$P$_{3/2}$ manifold.

\section{Challenges}

Several technical challenges must be met to successfully implement squeezing as described in the preceding
sections.  

\noindent\textit{Beam alignment} -
In order to strongly couple to the target center-of-mass mode, precise beam alignment must be
arranged. If the beams are offset to the side of the ion disc a tilt mode might be preferentially excited, or if they
push harder in the center than on the edges a drumhead mode might be excite more strongly. 

\noindent\textit{Off-resonant scattering} -
One of the most deleterious problems is that of decoherence of the superposition of spin states as a result of
off-resonant scattering of the drive beams. The scattering processes can broadly be classed as either \textit{Raman
scattering}, during which the atomic state after scattering differs from that before the scattering, and
\textit{Rayleigh scattering}, during which the atom returns to the same state after scattering.   To conserve
energy during Raman scattering, the scattered photon must have a slightly different frequency from the drive photon. As
a result a measurement on the photon will reveal ``which-path'' information and lead to strong decoherence.  Since
the photon energy does not change during a Rayleigh process, which-path information is not revealed in the same
way.  Consequently it has been commonly accepted that Raman scattering is the dominant decoherence process for
experiments with far detuned fields \cite{Leroux2010,Ozeri2007}, and that the contribution due to Rayleigh scattering is
negligible when the scattering rates are the same from the states $|\!\downarrow\rangle$ and $|\!\uparrow\rangle$.   

Recently it was shown that the decoherence effect due to Rayleigh scattering must calculated more carefully
\cite{Uys2010}. 
In particular, the Rayleigh decoherence of a spin-superposition depends on the square of the
difference between the sum of amplitudes for all scattering processes from the state $|\!\downarrow\rangle$ and the sum
of amplitudes for all scattering processes from the state $|\!\uparrow\rangle$.  Since the scattering amplitudes depend
on the detunings of the light fields, it is possible that for certain choices of detunings the decoherence contributions
due to scattering from
$|\!\downarrow\rangle$ might add constructively to the contribution from scattering from $|\!\uparrow\rangle$.  In fact,
even though
the drive fields might be far off-resonance, the Rayleigh contribution might be dominant over decoherence due to Raman
scattering. The laser detunings used in the Penning trap experiment fall in this regime.

\noindent\textit{Dephasing} - 
A secondary source of decoherence is dephasing, which results primarily from three sources: magnetic field fluctuations,
instability of the microwave phase reference and fluctuations in the AC Stark shift due to either power fluctuations in
the laser beam or beam pointing noise.  A single spin-echo $\pi$-pulse midway during the squeezing operation
mitigates the dephasing so that its effect can be neglected as compared to the decoherence resulting from spontaneous
light scattering.

\section{Conclusion}

Modern trends in metrology experiments indicate that measurement devices of the future will exploit pure quantum effects
to break old barriers in sensitivity.  Spin-squeezing is a promising quantum technique for surpassing shot-noise
limited measurements.  It is a technology currently being pursued by several laboratories world wide with promising
progress.  This proceedings discussed aspects of spin-squeezing using beryllium ions in a Penning ion trap which will
complement recent successes in neutral atom traps.\\

The Penning trap work is supported by the DARPA OLE program. NIST is an agency of the US government. This work is not
subject to US copyright.

\section{APPENDIX A}

We wish to construct properly normalized states generated
by$(J_-)^q|\uparrow\uparrow\uparrow...\uparrow\uparrow\rangle$. So
\begin{eqnarray}
(J_-)^q|\uparrow\uparrow\uparrow...\uparrow\uparrow\rangle&=&(\sum_j\hat\sigma^-_j)^q|\uparrow\uparrow\uparrow...
\uparrow\uparrow\rangle\\
&=&\sum_{j_1,j_2,...j_q}\hat\sigma^-_{j_1}\hat\sigma^-_{j_2}...\hat\sigma^-_{j_q}
|\uparrow\uparrow\uparrow...\uparrow\uparrow\rangle.
\label{A1}
\end{eqnarray}
Each index $j_k$ in \eqref{A1} runs from 1 to $N$.  However, whenever any two or more operators $\hat\sigma^-_{j_k}$
have the same index, the term will vanish after operation on the state
$|\uparrow\uparrow\uparrow...\uparrow\uparrow\rangle$, since a lowering operation cannot be applied to the
same particle twice. We therefore need only keep terms in the sum for which every index is different, and of which
there are $N(N-1)(N-2)...(N-(q-1))$, so
\begin{equation}
(J_-)^q|\uparrow\uparrow\uparrow...\uparrow\uparrow\rangle
=\sum\limits_{j_1\neq j_2\neq...\neq
j_q}\hat\sigma^-_{j_1}\hat\sigma^-_{j_2}...\hat\sigma^-_{j_q}|\uparrow\uparrow\uparrow...\uparrow\uparrow\rangle.
\label{A2}
\end{equation}
Now, after applying all operators $\hat\sigma^-_{j_q}$ there are $q!$ duplicates of each state\\
$|\uparrow\uparrow\downarrow_{j_1}...\downarrow_{j_2}...\downarrow_{j_q}\uparrow\rangle$.  For example, if $q=3$ 
each of the $3!$ permutations: $\hat\sigma^-_1\hat\sigma^-_2\hat\sigma^-_3$,
$\hat\sigma^-_1\hat\sigma^-_3\hat\sigma^-_2$, $\hat\sigma^-_2\hat\sigma^-_1\hat\sigma^-_3$, \textit{etc}.~leads to the
same state $|\downarrow\downarrow\downarrow\uparrow...\uparrow\rangle$. So keeping only unique terms we can rewrite
\eqref{A2} as
\begin{equation}
(J_-)^q|\uparrow\uparrow\uparrow...\uparrow\uparrow\rangle
=q!\sum\limits_{j_1<j_2<...<j_q}\hat\sigma^-_{j_1}\hat\sigma^-_{j_2}...\hat\sigma^-_{j_q}
|\uparrow\uparrow\uparrow...\uparrow\uparrow\rangle.
\label{A3}
\end{equation}
Notice in \eqref{A3} the change in indexing of the sum as compared to \eqref{A2} and the extra factor $q!$.  Equation
\eqref{A3} contains $\mathcal{N}=N(N-1)(N-2)...(N-(q-1))/q!$ terms so to normalize it we must divide by
$q!\sqrt{\mathcal{N}}=\sqrt{q!N(N-1)(N-2)...(N-(q-1))}$. Finally then, our properly normalized states are
\begin{eqnarray}
\hspace*{-1cm}|\frac{N}{2},\frac{N}{2}-q\rangle\!\!\!\!\!\!&=&\!\!\!\!\!\!\frac{1}{\sqrt{
q!N(N-1)...(N-(q-1)) } } (J_-)^q|\uparrow\uparrow\uparrow...\uparrow\uparrow\rangle\\
\hspace*{-1cm}\!\!\!\!\!\!&=&\!\!\!\!\!\!\frac{\sqrt{q!}}{\sqrt{N(N-1)...(N-(q-1))}}\sum\limits_{j_<j_2<...<j_q}
\hat\sigma^-_{j_1} \hat\sigma^-_{ j_2} ...\hat\sigma^-_{j_q }
|\uparrow\uparrow\uparrow...\uparrow\uparrow\rangle.
\label{A4}
\end{eqnarray}
We leave it to the reader to verify that these states will obey the expected raising and lowering relations of angular
momentum states:
\begin{equation}
 J_\pm |J,M_J\rangle = \sqrt{J(J+1)- M_J(M_J\pm 1)}|J,M_J\pm 1\rangle.
\end{equation}





\bibliography{spinsqueezebib}

\begin{thebibliography}{16}
\expandafter\ifx\csname natexlab\endcsname\relax\def\natexlab#1{#1}\fi
\expandafter\ifx\csname bibnamefont\endcsname\relax
  \def\bibnamefont#1{#1}\fi
\expandafter\ifx\csname bibfnamefont\endcsname\relax
  \def\bibfnamefont#1{#1}\fi
\expandafter\ifx\csname citenamefont\endcsname\relax
  \def\citenamefont#1{#1}\fi
\expandafter\ifx\csname url\endcsname\relax
  \def\url#1{\texttt{#1}}\fi
\expandafter\ifx\csname urlprefix\endcsname\relax\def\urlprefix{URL }\fi
\providecommand{\bibinfo}[2]{#2}
\providecommand{\eprint}[2][]{\url{#2}}

\bibitem[{\citenamefont{Chou et~al.}(2010)\citenamefont{Chou, Hume, Koelemeij,
  Wineland, and Rosenband}}]{Chou2010}
\bibinfo{author}{\bibfnamefont{C.}~\bibnamefont{Chou}},
  \bibinfo{author}{\bibfnamefont{D.}~\bibnamefont{Hume}},
  \bibinfo{author}{\bibfnamefont{J.}~\bibnamefont{Koelemeij}},
  \bibinfo{author}{\bibfnamefont{D.}~\bibnamefont{Wineland}}, \bibnamefont{and}
  \bibinfo{author}{\bibfnamefont{T.}~\bibnamefont{Rosenband}},
  \bibinfo{journal}{Phys. Rev. Lett.} \textbf{\bibinfo{volume}{7}},
  \bibinfo{pages}{070802} (\bibinfo{year}{2010}).

\bibitem[{\citenamefont{Combes and Wiseman}(2005)}]{CombesWiseman2005}
\bibinfo{author}{\bibfnamefont{J.}~\bibnamefont{Combes}} \bibnamefont{and}
  \bibinfo{author}{\bibfnamefont{H.}~\bibnamefont{Wiseman}},
  \bibinfo{journal}{J. Opt. B: Quantum Semiclass.}
  \textbf{\bibinfo{volume}{7}}, \bibinfo{pages}{14} (\bibinfo{year}{2005}).

\bibitem[{\citenamefont{Appel et~al.}(2009)\citenamefont{Appel, Windpassinger,
  Oblak, Hoff, Kjaergaard, and Polzik}}]{Appel2009}
\bibinfo{author}{\bibfnamefont{J.}~\bibnamefont{Appel}},
  \bibinfo{author}{\bibfnamefont{P.}~\bibnamefont{Windpassinger}},
  \bibinfo{author}{\bibfnamefont{D.}~\bibnamefont{Oblak}},
  \bibinfo{author}{\bibfnamefont{U.}~\bibnamefont{Hoff}},
  \bibinfo{author}{\bibfnamefont{N.}~\bibnamefont{Kjaergaard}},
  \bibnamefont{and} \bibinfo{author}{\bibfnamefont{E.}~\bibnamefont{Polzik}},
  \bibinfo{journal}{PNAS} \textbf{\bibinfo{volume}{106}},
  \bibinfo{pages}{10960} (\bibinfo{year}{2009}).

\bibitem[{\citenamefont{Takano et~al.}(2009)\citenamefont{Takano, Fuyama,
  Namiki, and Takahashi}}]{Takano2009}
\bibinfo{author}{\bibfnamefont{T.}~\bibnamefont{Takano}},
  \bibinfo{author}{\bibfnamefont{M.}~\bibnamefont{Fuyama}},
  \bibinfo{author}{\bibfnamefont{R.}~\bibnamefont{Namiki}}, \bibnamefont{and}
  \bibinfo{author}{\bibfnamefont{Y.}~\bibnamefont{Takahashi}},
  \bibinfo{journal}{Phys. Rev. Lett.} \textbf{\bibinfo{volume}{102}},
  \bibinfo{pages}{033601} (\bibinfo{year}{2009}).

\bibitem[{\citenamefont{Riedel et~al.}(2010)\citenamefont{Riedel, Bohi, Li,
  Hansch, Sinatra, and Treutlein}}]{Riedel2010}
\bibinfo{author}{\bibfnamefont{M.}~\bibnamefont{Riedel}},
  \bibinfo{author}{\bibfnamefont{P.}~\bibnamefont{Bohi}},
  \bibinfo{author}{\bibfnamefont{Y.}~\bibnamefont{Li}},
  \bibinfo{author}{\bibfnamefont{T.}~\bibnamefont{Hansch}},
  \bibinfo{author}{\bibfnamefont{A.}~\bibnamefont{Sinatra}}, \bibnamefont{and}
  \bibinfo{author}{\bibfnamefont{P.}~\bibnamefont{Treutlein}},
  \bibinfo{journal}{Nature} \textbf{\bibinfo{volume}{464}},
  \bibinfo{pages}{1170} (\bibinfo{year}{2010}).

\bibitem[{\citenamefont{Gross et~al.}(2010)\citenamefont{Gross, Zibold,
  Nicklas, Esteve, and Oberthaler}}]{Gross2010}
\bibinfo{author}{\bibfnamefont{C.}~\bibnamefont{Gross}},
  \bibinfo{author}{\bibfnamefont{T.}~\bibnamefont{Zibold}},
  \bibinfo{author}{\bibfnamefont{E.}~\bibnamefont{Nicklas}},
  \bibinfo{author}{\bibfnamefont{J.}~\bibnamefont{Esteve}}, \bibnamefont{and}
  \bibinfo{author}{\bibfnamefont{M.}~\bibnamefont{Oberthaler}},
  \bibinfo{journal}{Nature} \textbf{\bibinfo{volume}{464}},
  \bibinfo{pages}{1165} (\bibinfo{year}{2010}).

\bibitem[{\citenamefont{Leroux et~al.}(2010)\citenamefont{Leroux,
  Schleier-Smith, and Vuletic}}]{Leroux2010}
\bibinfo{author}{\bibfnamefont{I.}~\bibnamefont{Leroux}},
  \bibinfo{author}{\bibfnamefont{M.}~\bibnamefont{Schleier-Smith}},
  \bibnamefont{and} \bibinfo{author}{\bibfnamefont{V.}~\bibnamefont{Vuletic}},
  \bibinfo{journal}{Phys. Rev. Lett.} \textbf{\bibinfo{volume}{104}},
  \bibinfo{pages}{073602} (\bibinfo{year}{2010}).

\bibitem[{\citenamefont{Chen et~al.}(2011)\citenamefont{Chen, Bohnet, Sankar,
  and Thompson}}]{ChenBohnet2011}
\bibinfo{author}{\bibfnamefont{Z.}~\bibnamefont{Chen}},
  \bibinfo{author}{\bibfnamefont{J.}~\bibnamefont{Bohnet}},
  \bibinfo{author}{\bibfnamefont{S.}~\bibnamefont{Sankar}}, \bibnamefont{and}
  \bibinfo{author}{\bibfnamefont{J.}~\bibnamefont{Thompson}},
  \bibinfo{journal}{Phys. Rev. Lett.} \textbf{\bibinfo{volume}{106}},
  \bibinfo{pages}{133601} (\bibinfo{year}{2011}).

\bibitem[{\citenamefont{Kitagawa and Ueda}(1993)}]{Kitagawa1993}
\bibinfo{author}{\bibfnamefont{M.}~\bibnamefont{Kitagawa}} \bibnamefont{and}
  \bibinfo{author}{\bibfnamefont{M.}~\bibnamefont{Ueda}},
  \bibinfo{journal}{Phys. Rev. A} \textbf{\bibinfo{volume}{47}},
  \bibinfo{pages}{5138} (\bibinfo{year}{1993}).

\bibitem[{\citenamefont{Meyer et~al.}(2001)\citenamefont{Meyer, Rowe,
  Kielpinski, Sackett, Itano, Monroe, and Wineland}}]{MeyerRowe2001}
\bibinfo{author}{\bibfnamefont{V.}~\bibnamefont{Meyer}},
  \bibinfo{author}{\bibfnamefont{M.}~\bibnamefont{Rowe}},
  \bibinfo{author}{\bibfnamefont{D.}~\bibnamefont{Kielpinski}},
  \bibinfo{author}{\bibfnamefont{C.}~\bibnamefont{Sackett}},
  \bibinfo{author}{\bibfnamefont{W.}~\bibnamefont{Itano}},
  \bibinfo{author}{\bibfnamefont{C.}~\bibnamefont{Monroe}}, \bibnamefont{and}
  \bibinfo{author}{\bibfnamefont{D.}~\bibnamefont{Wineland}},
  \bibinfo{journal}{Phys. Rev. Let.} \textbf{\bibinfo{volume}{86}},
  \bibinfo{pages}{5870} (\bibinfo{year}{2001}).

\bibitem[{\citenamefont{Dicke}(1954)}]{Dicke1954}
\bibinfo{author}{\bibfnamefont{R.}~\bibnamefont{Dicke}},
  \bibinfo{journal}{Phys. Rev.} \textbf{\bibinfo{volume}{93}},
  \bibinfo{pages}{99} (\bibinfo{year}{1954}).

\bibitem[{\citenamefont{Leibfried et~al.}(2003)\citenamefont{Leibfried,
  DeMarco, Meyer, Lucas, Barret, Britton, Itano, Jelenkovic, Langer, Rosenband
  et~al.}}]{Leibfried2003}
\bibinfo{author}{\bibfnamefont{D.}~\bibnamefont{Leibfried}},
  \bibinfo{author}{\bibfnamefont{B.}~\bibnamefont{DeMarco}},
  \bibinfo{author}{\bibfnamefont{V.}~\bibnamefont{Meyer}},
  \bibinfo{author}{\bibfnamefont{D.}~\bibnamefont{Lucas}},
  \bibinfo{author}{\bibfnamefont{M.}~\bibnamefont{Barret}},
  \bibinfo{author}{\bibfnamefont{J.}~\bibnamefont{Britton}},
  \bibinfo{author}{\bibfnamefont{W.}~\bibnamefont{Itano}},
  \bibinfo{author}{\bibfnamefont{B.}~\bibnamefont{Jelenkovic}},
  \bibinfo{author}{\bibfnamefont{C.}~\bibnamefont{Langer}},
  \bibinfo{author}{\bibfnamefont{T.}~\bibnamefont{Rosenband}},
  \bibnamefont{et~al.}, \bibinfo{journal}{Nature}
  \textbf{\bibinfo{volume}{422}}, \bibinfo{pages}{412} (\bibinfo{year}{2003}).

\bibitem[{\citenamefont{Mitchell et~al.}(1998)\citenamefont{Mitchell,
  Bollinger, Dubin, Huang, Itano, and Baughman}}]{MitchellBollinger1998}
\bibinfo{author}{\bibfnamefont{T.}~\bibnamefont{Mitchell}},
  \bibinfo{author}{\bibfnamefont{J.}~\bibnamefont{Bollinger}},
  \bibinfo{author}{\bibfnamefont{D.}~\bibnamefont{Dubin}},
  \bibinfo{author}{\bibfnamefont{X.}~\bibnamefont{Huang}},
  \bibinfo{author}{\bibfnamefont{W.}~\bibnamefont{Itano}}, \bibnamefont{and}
  \bibinfo{author}{\bibfnamefont{R.}~\bibnamefont{Baughman}},
  \bibinfo{journal}{Science} \textbf{\bibinfo{volume}{282}},
  \bibinfo{pages}{1290} (\bibinfo{year}{1998}).

\bibitem[{\citenamefont{Biercuk et~al.}(2009)\citenamefont{Biercuk, Uys,
  Van~Devender, Shiga, Itano, and Bollinger}}]{Biercuk2009}
\bibinfo{author}{\bibfnamefont{M.}~\bibnamefont{Biercuk}},
  \bibinfo{author}{\bibfnamefont{H.}~\bibnamefont{Uys}},
  \bibinfo{author}{\bibfnamefont{A.}~\bibnamefont{Van~Devender}},
  \bibinfo{author}{\bibfnamefont{N.}~\bibnamefont{Shiga}},
  \bibinfo{author}{\bibfnamefont{W.}~\bibnamefont{Itano}}, \bibnamefont{and}
  \bibinfo{author}{\bibfnamefont{J.}~\bibnamefont{Bollinger}},
  \bibinfo{journal}{Quant. Inf. Comp.} \textbf{\bibinfo{volume}{9}},
  \bibinfo{pages}{0920} (\bibinfo{year}{2009}).

\bibitem[{\citenamefont{Ozeri et~al.}(2007)\citenamefont{Ozeri, Itano,
  Blakestad, Britton, Chiaverini, Jost, Langer, Leibfried, Reichle, Seidelin
  et~al.}}]{Ozeri2007}
\bibinfo{author}{\bibfnamefont{R.}~\bibnamefont{Ozeri}},
  \bibinfo{author}{\bibfnamefont{W.}~\bibnamefont{Itano}},
  \bibinfo{author}{\bibfnamefont{R.}~\bibnamefont{Blakestad}},
  \bibinfo{author}{\bibfnamefont{J.}~\bibnamefont{Britton}},
  \bibinfo{author}{\bibfnamefont{J.}~\bibnamefont{Chiaverini}},
  \bibinfo{author}{\bibfnamefont{J.}~\bibnamefont{Jost}},
  \bibinfo{author}{\bibfnamefont{C.}~\bibnamefont{Langer}},
  \bibinfo{author}{\bibfnamefont{D.}~\bibnamefont{Leibfried}},
  \bibinfo{author}{\bibfnamefont{R.}~\bibnamefont{Reichle}},
  \bibinfo{author}{\bibfnamefont{S.}~\bibnamefont{Seidelin}},
  \bibnamefont{et~al.}, \bibinfo{journal}{Phys. Rev. A}
  \textbf{\bibinfo{volume}{75}}, \bibinfo{pages}{042329}
  (\bibinfo{year}{2007}).

\bibitem[{\citenamefont{Uys et~al.}(2010)\citenamefont{Uys, Biercuk,
  VanDevender, Ospelkaus, Meiser, Ozeri, and Bollinger}}]{Uys2010}
\bibinfo{author}{\bibfnamefont{H.}~\bibnamefont{Uys}},
  \bibinfo{author}{\bibfnamefont{M.}~\bibnamefont{Biercuk}},
  \bibinfo{author}{\bibfnamefont{A.}~\bibnamefont{VanDevender}},
  \bibinfo{author}{\bibfnamefont{C.}~\bibnamefont{Ospelkaus}},
  \bibinfo{author}{\bibfnamefont{D.}~\bibnamefont{Meiser}},
  \bibinfo{author}{\bibfnamefont{R.}~\bibnamefont{Ozeri}}, \bibnamefont{and}
  \bibinfo{author}{\bibfnamefont{J.}~\bibnamefont{Bollinger}},
  \bibinfo{journal}{Phys. Rev. Lett.} \textbf{\bibinfo{volume}{105}},
  \bibinfo{pages}{200401} (\bibinfo{year}{2010}).

\end{thebibliography}

\end{document}